\documentclass[11pt]{article}

\usepackage[english]{babel}

\usepackage[T1]{fontenc}
\usepackage[symbol]{footmisc} % For footnote symbols.
\usepackage{mathptmx}
\usepackage{pifont} % For ding symbols.

\usepackage[affil-it]{authblk}
\usepackage{dirtytalk} % For quotes.
\usepackage{fancyhdr}
\usepackage{newfloat}
\usepackage[width=14cm]{geometry}
\usepackage{xspace}

\usepackage{amsmath}
\usepackage{amssymb}
\usepackage{mathtools}

\usepackage{enumitem} % To itemize.
\setlist[itemize]{
	itemsep=-2pt,
	topsep=2pt,
	leftmargin=3.5em,
	label=\ding{226}
}
\setlist[enumerate]{
	itemsep=-2pt,
	topsep=2pt,
	leftmargin=3.5em
}

\usepackage{listings}
\usepackage{xcolor}

\DeclareFloatingEnvironment[
	fileext=lol,
	placement=htbp
]{listing}

\definecolor{commentcolor}{rgb}{0.4,0.4,0.4}
\definecolor{identifiercolor}{rgb}{0.1,0.1,0.35}
\definecolor{keywordcolor}{rgb}{0,0.55,0.65}

\lstdefinestyle{mystyle}{
	basicstyle=\footnotesize\ttfamily,
	commentstyle=\itshape\color{commentcolor},
	identifierstyle=\color{identifiercolor},
	keywordstyle=\color{keywordcolor},
	language=C++,
	numbers=none,
	tabsize=4,
	columns=fullflexible,
	keepspaces=true
}
\usepackage[
labelfont=bf,
textfont=it,
skip=15pt
]{caption}
\usepackage{graphicx}
\usepackage{tabularray}
\usepackage{tikz}

\usetikzlibrary{arrows, calc, positioning, shapes}

\makeatletter
\g@addto@macro\@floatboxreset\centering
\makeatother

\DeclareTblrTemplate{caption-tag}{default}{}
\DeclareTblrTemplate{caption-sep}{default}{}
\DeclareTblrTemplate{caption-text}{default}
{\textbf{Table\hspace{0.25em}\thetable:}
	\hspace{-6pt}{\textit{\enskip\InsertTblrText{caption}}}
}

\DefTblrTemplate{firsthead}{default}{}
\DefTblrTemplate{lastfoot}{default}{
	\UseTblrTemplate{note}{default}
	\UseTblrTemplate{remark}{default}
	\vspace*{10pt}
	\UseTblrTemplate{caption}{default}
}

\DeclareTblrTemplate{capcont}{default}{
	\UseTblrTemplate{caption-tag}{default}
	\UseTblrTemplate{conthead-text}{default}
}
\SetTblrStyle{capcont}{font=\footnotesize}
\SetTblrStyle{contfoot}{font=\footnotesize}

\SetTblrStyle{note-text}{font=\footnotesize}

\usepackage[
backend=bibtex,
sorting=none,
sortcites=true,
style=numeric-comp,
isbn=false,
autolang=hyphen
]{biblatex}

\AtEveryBibitem{\clearlist{language}} % Remove language info.
\AtEveryBibitem{
	\clearfield{urlyear}
	\clearfield{urlmonth}
}

\DefineBibliographyStrings{english}{
	url = Available,
} % Remove '[Online]'.

\usepackage[linktoc=all]{hyperref}
\usepackage[capitalise,nameinlink]{cleveref}
\hypersetup{
	colorlinks=true, % Colour the text of links.
	urlcolor=blue, % Colour for \url and \href links.
	linkcolor=blue, % Colour for \nameref links.
	citecolor=blue, % Colour of reference citations.
}

\newcommand{\eA}{\textit{et al.}\xspace}
\newcommand{\iE}{\textit{i.e.}\xspace}
\newcommand{\superR}
	{\textsuperscript{\scriptsize{\textregistered}}\xspace}
\newcommand{\superTM}
	{\textsuperscript{\scriptsize{\texttrademark}}\xspace}

\newcommand{\config}[2]{\textit{#1\_#2}}
\newcommand{\isay}[1]{\say{\textit{#1}}}
\newcommand{\valms}[2]{#1$\pm$#2~ms}
\newcommand{\valperc}[2]{#1$\pm$#2\%}

\newcommand{\ETI}{\Delta T_I}
\newcommand{\ETP}{\Delta T_P}
\newcommand{\ETR}{\tau}
\newcommand{\stdf}{\sigma}

\newcommand{\code}[1]{\textnormal{\protect\lstinline{#1}}}

\def\tablelinewidth{1pt}
\def\tablestretch{1.2}

\title{AVX / NEON Intrinsic Functions: \\ When Should They Be Used?
	\footnote{
		\includegraphics[height=8pt,trim=0 2pt 0 0]{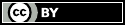} Distributed 		
		under a 
		\href{https://creativecommons.org/licenses/by/4.0/}{CC-BY 4.0 licence}.
	}
}

\author[]{
	Théo Boivin
	\hspace{-5pt}
	\footnote{
		Email:
		\href{mailto:theo.boivin@email.fr}{theo.boivin@email.fr}
	}
	\hspace{-2pt}
}

\author[]{
	\hspace{-5pt}
	Joeffrey Legaux
	\hspace{-5pt}
	\footnote{
		Email:
		\href{mailto:joeffrey.legaux@cerfacs.fr}{joeffrey.legaux@cerfacs.fr}
	}
}

\affil[]{CERFACS, 31057 Toulouse Cedex 1, France}

\date{}

\begin{document}

\maketitle

\vspace{-0.2cm}

\begin{abstract}
	A cross-configuration benchmark is proposed to explore the capacities and
	limitations of AVX / NEON intrinsic functions in a generic context of
	development project, when a vectorisation strategy is required to optimise 
	the	code. The main aim is to guide developers to choose when using 
	intrinsic functions, depending on the OS, architecture and/or available 
	compiler. Intrinsic functions were observed highly efficient in conditional 
	branching, with intrinsic version execution time reaching around 5\% of 
	plain code execution time. However, intrinsic functions were observed as 
	unnecessary in many cases, as the compilers already well auto-vectorise the 
	code.
\end{abstract}

\vspace{1cm}

\hspace{8pt}
\includegraphics[width=0.86\textwidth]{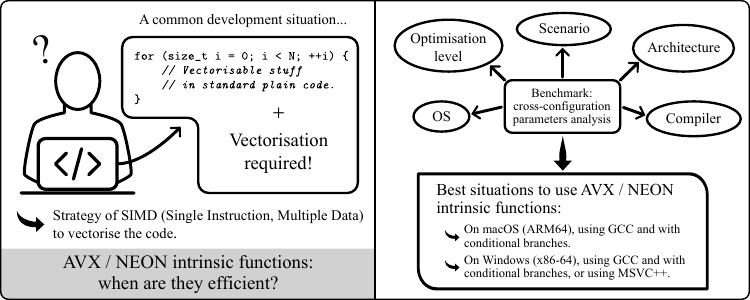}

\vspace*{38pt}

% Keywords
%  - AVX
%  - NEON
%  - Intrinsic functions
%  - Guidelines
%  - Performance
%  - Benchmarking

\pagebreak

\section{Introduction}

When it comes to programming, optimisation is at the core of computation
efficiency. If the computation speed need is real, which is a fundamental
prerequisite to talk about optimisation, a possible technique is the SIMD
(Single Instruction, Multiple Data) that consists in applying a single
instruction to multiple data simultaneously, on a single hardware unit. In this
case, the data is said \say{vectorised}.

The CPU (Central Processing Unit), and more precisely a core of a CPU on current
common architectures such as x86-64 or ARM, is commonly seen as a scalar device:
its instructions only process a single piece of data at a time. However, this
description is not exact. The instructions are applied on a register, a memory
of limited capacity going from 64 to 512 bits depending on the chip. The
manufacturers worked to extend the usage of these registers, introducing in the
late 1990s the first SIMD instruction sets to the x86 architecture instruction
set, first the MMX introduced by Intel in 1997, then \textit{3DNow!} proposed by
AMD in 1998, then the SSE (Streaming SIMD Extensions) introduced by Intel in
1999. The SSE instruction set originally used eight 128-bits registers called
XMM0-XMM7. This instruction set made possible the vectorisation of floating
operations, more precisely four 32-bits single-precision floating-points or two
64-bits double-precision floating-points. At the second generation of
Intel\superR Core\superTM processor family (nicknamed \say{Sandy Bridge}) in
2011, Intel introduced sixteen 256-bit registers called YMM0-YMM15 and the AVX
(Advanced Vector Extensions) instruction set, leading to the doubling of
registers memory. This memory was doubled once again in 2013 through the
introduction of thirty-two 512-bits registers called ZMM0-ZMM31 within the
Intel\superR Xeon Phi\superTM processor (nicknamed \say{Knights Landing}), as
well as the AVX-512 instruction set that introduced numerous features (masking,
enhanced mathematics, better standard language compliance)
\cite{gepner_early_2011, anderson_enhanced_2018, mileff_improving_2022,
intel_AVX-512_2017, intel_instruction_2025}.

Nowadays, the intrinsic functions (called \say{intrinsics} in this paper)
belongs to the toolbox of developers to explicitly manipulate vector
instruction sets and improve the vectorisation of computation, mostly in C++.
Many researches demonstrate the enhanced performances using intrinsics. For
example, Jeong \eA (2012) \cite{jeong_performance_2012} compared the plain,
SSE4.2, AVX1 intrinsic versions of a simple addition loop, where the term
\say{plain} refers to the C++ standard version of a code, without any use of
intrinsic. They measured clear improvement by adjusting the data reuse,
decreasing the required number of CPU clocks by around 11 with AVX version.
Jinchen \eA (2012) \cite{jinchen_optimization_2012} leveraged data reuse to
optimise mathematical functions (\code{cos}, \code{sin}, \code{sqrt}, etc.),
obtaining an average improvement of around 8\%. Hassan \eA (2016, 2018)
\cite{hassan_performance_2016, hassan_effective_2018} addressed the algorithm of
large matrices multiplication. They detected a better performance of Intel
compiler compared to MSVC++ and an improvement with their proposed algorithm
between 14\% and 18\%. Shabanov \eA (2019) \cite{shabanov_vectorization_2019}
studied the application of AVX-512 intrinsics in conditional programming. As an
important note, they advised to remove the unlikely conditional branches from
the main context, on one hand to facilitate vectorisation, and on the other hand
to avoid unnecessary vectorised computation. Cebrian \eA (2020)
\cite{cebrian_scalability_2020} mostly focused on energy consumption by
comparing thread-level parallelisation and SIMD. They noticed that even though
both approaches were comparable in terms of computation speed, SIMD consumed
much less power than threading. Fortin \eA (2021)
\cite{fortin_highperformance_2021} optimised the algorithms of polynomial
factorisation and polynomial greatest common divisor with two versions,
respectively using AVX2 and AVX-512 intrinsics. The first version gets a speedup
factor of 3.7 and the second one a speedup factor of 7.2. In addition to these
raw performance studies, vectorisation using AVX intrinsics was used to improve
the performance of many applications, such as fluid vibration problems
\cite{frances_performance_2013}, heart simulation \cite{jarvis_combining_2019},
N-body problem \cite{pedregosa-gutierrez_direct_2021}, seismic wave propagation
\cite{jubertie_portability_2021} or particle swarm simulation
\cite{safarik_acceleration_2023}.

The current literature consequently underlines the efficiency of optimisation
using AVX intrinsics. The quick review presented here focused on AVX instruction
set, adapted for x86-64 architecture, but the equivalent instruction set for ARM
architecture, NEON, can be joined to the following observation: using
intrinsics, whether AVX or NEON, seems quite underestimated in general
development practice. The original question at the origin of this paper mostly
raised from this general under-use: the optimisation of CPU vectorisation
through programming is not particularly common. Consequently, for a developer
discovering AVX / NEON intrinsics capabilities, the task is hard to answer a
simple question: \textit{For my project, are AVX / NEON intrinsics interesting?}
Indeed, most of cited references are mostly punctual in terms of configuration
(OS, CPU architecture, compiler). In this sense, \cref{tab:review} lists all the
configurations of all cited references. The majority of experiments were carried
out on x86-64 architecture, using ICC (Intel C++ Compiler) or GCC (GNU Compiler
Collection) compilers. Some, however, made a comparison between two compilers
\cite{hassan_performance_2016, hassan_effective_2018, jarvis_combining_2019,
fortin_highperformance_2021, jubertie_portability_2021}. The work presented in
these papers remain quite impressive and complete, but a lack exists concerning
the wider guidelines perspective, essential to orientate developers to use or
not AVX / NEON intrinsics in their project. The question of \isay{When
intrinsics should be used?} is legitimate, because an aspect that few references
underline is their cost in terms of readability: intrinsics are tedious.
Readability and maintainability in a development project are as important as
efficiency, so intrinsics should be chosen wisely. Furthermore, the
implementation of intrinsics is not systematically efficient. For example,
Gottschlag \eA (2020a and 2020b) \cite{gottschlag_automatic_2020,
gottschlag_avx_2020} questioned the efficiency of AVX instructions. They studied
the CPU frequency reduction generated by the processor during AVX-512 intrinsics
to avoid over-consumption. The delay of frequency increase after the AVX-512
code execution may overlap on plain code execution, which reduces the efficiency
of the latter. This leads to a balance between AVX-512 code speedup and
surrounding plain code speedown, only detectable within the main program. This
constitutes a supplementary reason to think intrinsics carefully, in order to
use them optimally.

\pagebreak

\vspace*{5pt}
\begin{longtblr}[
	baseline=t,
	caption = {Review of cited references configurations
		---
		An empty cell means that no data was provided / found in the reference.
		\textbf{GCC}: GNU Compiler Collection,
		\textbf{ICC}: Intel C++ Compiler,
		\textbf{MSVC++}: Microsoft Visual C++.
	},
	note{$\dag$} = {Not clearly explicated in reference, deduced from content.},
	label = {tab:review}
	]{
		colspec={
			Q[c,m,4cm]
			Q[c,m,3cm]
			Q[c,m,1.8cm]
			Q[c,m,3.4cm]},
		rows={font=\footnotesize},
		row{1}={abovesep+=2pt, font=\normalsize},
		hline{1,2,17}={1-Z}{\tablelinewidth},
		stretch=\tablestretch
	}
	Reference & OS / Kernel & Architecture & Compiler \\
	
	Gepner \eA \cite{gepner_early_2011} &
	RedHat Enterprise Linux 6, kernel 2.6.3271.el6.x86\_64 &
	x86-64 &
	ICC v12.0 \\
	
	Anderson \eA \cite{anderson_enhanced_2018} &
	-- &
	x86-64 &
	-- \\
	
	Mileff \eA \cite{mileff_improving_2022} &
	Linux &
	x86-64 &
	GCC v11.1 \\
	
	Jeong \eA \cite{jeong_performance_2012} &
	{Fedora  17, \\ kernel 3.5.2-3.fc17} &
	x86-64 &
	GCC v4.7 \\
	
	Jinchen \eA \cite{jinchen_optimization_2012} &
	-- &
	-- &
	-- \\
	
	Hassan \eA \cite{hassan_performance_2016} &
	Windows 10 &
	x86-64 &
	{ICC / \\ MSVC++ 2015 v140} \\
	
	Hassan \eA \cite{hassan_effective_2018} &
	Windows 10 &
	x86-64 &
	{ICC v17.0 / \\ MSVC++ 2015} \\
	
	Shabanov \eA \cite{shabanov_vectorization_2019} &
	-- &
	x86-64 &
	-- \\
	
	Cebrian \eA \cite{cebrian_scalability_2020} &
	{Ubuntu 18.04, \\ kernel 4.15} &
	x86-64 &
	GCC v7.3 \\
	
	Fortin \eA \cite{fortin_highperformance_2021} &
	-- &
	x86-64 &
	{GCC v8.2 / \\ ICC v19.0} \\
	
	Francés \eA \cite{frances_performance_2013} &
	-- &
	x86-64 &
	GCC \TblrNote{$\dag$} \\
	
	Jarvis \eA \cite{jarvis_combining_2019} &
	-- &
	x86-64 &
	{GCC v8.2 / \\ ICC v18.0} \\
	
	Pedregosa-Gutierrez \eA \cite{pedregosa-gutierrez_direct_2021} &
	-- &
	x86-64 &
	ICC \TblrNote{$\dag$} \\
	
	Jubertie \eA \cite{jubertie_portability_2021} &
	-- &
	ARM &
	{Armclang v20.0 / \\ GCC v10.0} \\
	
	Safarik \& Snasel \cite{safarik_acceleration_2023} &
	Windows 10 &
	x86-64 &
	-- \\
\end{longtblr}

\pagebreak

The aim of this paper is to propose a cross-configuration benchmark, based on
simple generic tests, in order to estimate the configurations where AVX / NEON
intrinsics are susceptible to notably improve performance. An important
assumption concerning the development project is stated here: vectorisation is
the chosen strategy. The aim of this paper is not to evaluate the efficiency of
vectorised code compared to non-vectorised code, but instead, to compare
auto-vectorised code and manually vectorised code. Even if this paper cannot
pretend to be exhaustive for all development situations, the main idea is to
provide a wider perspective on AVX / NEON intrinsics capabilities and
limitations, in order to draw a clearer routine of choice for intrinsics use.

\section{Experiment}

\subsection{Scenarios}

The experiment consisted in several sample programs commonly met in development
practice. When it comes to SIMD, the vectorised program must present a
loop-like form, \iE implements a common instruction on multiple data. Thus, each
sample program was a loop of operations made on standard vectors. Several
scenarios were approached, starting from basic operations (addition,
multiplication) to more and more complex cases, using index offset, advanced
operations (\code{cos}, \code{sqrt}, \code{pow}, etc.), conditions on index and
conditions on random data. In total, eight scenarios were benchmarked,
detailed in \cref{list:plain-scenarios} with descriptions and plain codes. Most
of the additional complexity\footnote[2]{\hspace{3pt}The term \say{complexity}
refers to the problem complexity, not the computational algorithm complexity.}
implemented at each scenario (for example the index offset or the conditions)
were chosen to test the auto-vectorisation of compilers. The intrinsic
implementation was made using \say{SIMD Everywhere} \cite{github_simde}, an
open-source header-only library that provides fast and portable intrinsic
implementations, for both AVX and NEON instruction sets. In order to ensure a
fully portable code between devices, only 256-bit registers instructions were
used (\code{simde__m256} type). The implementation strategy was based on the
load / store technique: loading data in intermediary variables through
\code{simde_mm256_loadu_ps()} method, computing, then storing back in standard
vectors using \code{simde_mm256_storeu_ps()}. For advanced operations
(\code{cos}, \code{sqrt}, \code{pow}, etc.), the intrinsic functions were used
(\code{simde_mm256_cos_ps()}, \code{simde_mm256_sqrt_ps()},
\code{simde_mm256_pow_ps()}, etc.). As a set of examples,
\cref{list:intrinsic-scenarios} details the intrinsic versions of scenarios 1
and 6. Vectors \code{A}, \code{B}, \code{C} and \code{D} are written with the
same names in both \cref{list:plain-scenarios} and
\cref{list:intrinsic-scenarios} for clarity, but they were distinguished in the
benchmark to avoid the influence of memory cache.

\begin{listing}
	\begin{tabular}{|p{0.44\textwidth}|p{0.48\textwidth}|}
		\vspace{-1em}
		\begin{minipage}[t]{0.5\textwidth}
			\begin{lstlisting}
// Scenario 1
// Basic operations.
for (size_t i = 0; i < Niter; ++i) {
	D[i] = A[i] * B[i] + C[i];
}
		\end{lstlisting}
		\end{minipage} &
		\vspace{-1em}
		\begin{minipage}[t]{0.5\textwidth}
			\begin{lstlisting}
// Scenario 2
// Basic operations with index offset.
for (size_t i = 1; i < Niter-1; ++i) {
	D[i] = A[i-1] * B[i] + C[i] + B[i+1];
}
D[0] = 0.0f;
D[Niter-1] = 0.0f;
			\end{lstlisting}
		\end{minipage} \\
		\vspace*{-2.5em}
		\begin{minipage}[t]{0.5\textwidth}
			\begin{lstlisting}
// Scenario 3
// Advanced operations.
//
for (size_t i = 0; i < Niter; ++i) {
	D[i] = A[i] * std::sqrt(B[i])
		+ std::abs(C[i])
		- std::cos(A[i]) / C[i]
		+ std::pow(B[i], 2.5f);
}
			\end{lstlisting}
		\end{minipage} &
		\vspace*{-2.5em}
		\begin{minipage}[t]{0.5\textwidth}
			\begin{lstlisting}
// Scenario 4
// Advanced operations with index
// offset.
for (size_t i = 1; i < Niter-1; ++i) {
	D[i] = A[i-1] * std::sqrt(B[i])
		+ std::abs(C[i])
		- std::cos(A[i]) / C[i]
		+ std::pow(B[i+1], 2.5f);
}
D[0] = 0.0f;
D[Niter-1] = 0.0f;
			\end{lstlisting}
		\end{minipage} \\
		\vspace*{-2.5em}
		\begin{minipage}[t]{0.5\textwidth}
			\begin{lstlisting}
// Scenario 5
// Simple condition on index
// with basic operations.
for (size_t i = 0; i < Niter; ++i) {
	if (i % 2 == 0)
		C[i] = A[i] + B[i];
	else
		C[i] = A[i] - B[i];
}
			\end{lstlisting}
		\end{minipage} &
		\vspace*{-2.5em}
		\begin{minipage}[t]{0.5\textwidth}
			\begin{lstlisting}
// Scenario 6
// Simple condition on random data
// with basic operations.
for (size_t i = 0; i < Niter; ++i) {
	if (A[i] > 5)
		C[i] = A[i] + B[i];
	else
		C[i] = A[i] - B[i];
}
			\end{lstlisting}
		\end{minipage} \\
		\vspace*{-2.5em}
		\begin{minipage}[t]{0.5\textwidth}
			\begin{lstlisting}
// Scenario 7
// Simple condition on random data
// with sub-branches and basic 
// operations.
for (size_t i = 0; i < Niter; ++i) {
	if (A[i] > 5) {
		if (B[i] >= 8)
			C[i] = A[i] * B[i];
		else if (B[i] <= 5)
			C[i] = A[i] / B[i];
		else
			C[i] = A[i] + B[i];
	}
	else
		C[i] = A[i] - B[i];
}
			\end{lstlisting}
		\end{minipage}
		\vspace*{-1.8em}
		&
		\vspace*{-2.5em}
		\begin{minipage}[t]{0.5\textwidth}
			\begin{lstlisting}
// Scenario 8
// Simple condition on random data
// with sub-branches and advanced
// operations.
for (size_t i = 0; i < Niter; ++i) {
	if (A[i] > 5) {
		if (B[i] >= 8)
			C[i] = std::sqrt(A[i]);
		else if (B[i] <= 5)
			C[i] = std::pow(A[i], B[i]);
		else
			C[i] = std::cos(A[i]);
	}
	else
		C[i] = std::ceil(A[i]);
}
			\end{lstlisting}
		\end{minipage}
		\vspace*{-1.8em}
	\end{tabular}
	\caption{Scenarios plain codes
		---
		Each scenario consists of a loop over \code{Niter = 5e7}, on
		\code{std::vector<float>} \code{A, B, C, D}. For scenario 6, vector 
		\code{A} was filled with random floats between 1 and 10. For scenarios 
		7 and 8, both vectors \code{A} and \code{B} were filled with random 
		floats between 1 and 10.
	}
	\label{list:plain-scenarios}
\end{listing}

\begin{listing}
	\begin{tabular}{|p{0.95\textwidth}|}
		\vspace{-1em}
		\begin{lstlisting}
// Scenario 1
simde__m256 a;
simde__m256 b;
simde__m256 c;
simde__m256 d;
for (size_t i = 0; i <= Niter - AVX_PS_VEC_SIZE; i += AVX_PS_VEC_SIZE) {
	a = simde_mm256_loadu_ps(&A[i]);
	b = simde_mm256_loadu_ps(&B[i]);
	c = simde_mm256_loadu_ps(&C[i]);
	d = simde_mm256_mul_ps(a, b);
	d = simde_mm256_add_ps(d, c);
	simde_mm256_storeu_ps(&D[i], d);
}
for (size_t i = Niter - Niter % AVX_PS_VEC_SIZE; i < Niter; ++i) {
	D[i] = A[i] * B[i] + C[i];
}
		\end{lstlisting}
		\vspace*{-1em}
		\begin{lstlisting}
// Scenario 6
simde__m256 a;
simde__m256 b;
simde__m256 c;
simde__m256 mask;
simde__m256 val_f_5 = simde_mm256_set1_ps(5.0f); 
for (size_t i = 0; i <= Niter - AVX_PS_VEC_SIZE; i += AVX_PS_VEC_SIZE) {
	a = simde_mm256_loadu_ps(&A[i]);
	b = simde_mm256_loadu_ps(&B[i]);
	mask = simde_mm256_cmp_ps(a, val_f_5, SIMDE_CMP_GT_OS);
	c = simde_mm256_blendv_ps(
		simde_mm256_sub_ps(a, b),
		simde_mm256_add_ps(a, b),
		mask
	);
	simde_mm256_storeu_ps(&C[i], c);
}
for (size_t i = Niter - Niter % AVX_PS_VEC_SIZE; i < Niter; ++i) {
	if (A[i] > 5)
		C[i] = A[i] + B[i];
	else
		C[i] = A[i] - B[i];
}
		\end{lstlisting}
		\vspace*{-1.25em}
	\end{tabular}
	\caption{Scenarios intrinsic codes (1 and 6)
		---
		Each scenario consists of a loop over \code{Niter - AVX_PS_VEC_SIZE}, 
		with step \code{AVX_PS_VEC_SIZE}, on \code{std::vector<float>} \code{A, 
		B, C, D}. We have \code{Niter = 5e7}, \code{AVX_PS_VEC_SIZE} the vector 
		size with floats contained in 256-bit register: \code{constexpr size_t} 
		\code{AVX_PS_VEC_SIZE} \code{=} \code{256 / (8 * sizeof(float))}. 
		Vectors \code{A, B, C, D} did not share the same memory as vectors 
		\code{A, B, C, D} in \cref{list:plain-scenarios}.
	}
	\label{list:intrinsic-scenarios}
\end{listing}

\newpage

\subsection{Configurations}

Three different devices were used, one for each main operating system (Linux,
macOS and Windows). For each operating system, one or more different compilers
were tested among Clang, GCC, ICC and MSVC++. The compiler GCC has been chosen
as the reference one (tested for all operating systems), using a common version
(14.3.0). In total, six configurations were tested: GCC on Linux, Clang and GCC
on macOS, GCC, ICC and MSVC++ on Windows, respectively named \config{lin}{gcc},
\config{mac}{clang}, \config{mac}{gcc}, \config{win}{gcc}, \config{win}{intel}
and \config{win}{msvc}. For further details, \cref{tab:configurations} details
the hardware / software configurations. For each compiler, the level of
optimisation was also addressed by testing the most common levels (O0, O1, O2
and O3). For ICC and MSVC++ compilers, the equivalent level for O0 was Od. For
MSVC++ compiler, the level O3 was not tested as, the maximum is O2.

\begin{table}[b!]
	\begin{tblr}{
			colspec={
				Q[c,m,1.5cm]
				Q[c,m,4cm]
				Q[c,m,2cm]
				Q[c,m,1cm]
				Q[c,m,1.5cm]},
			rows={font=\footnotesize},
			row{1}={abovesep+=2pt, font=\normalsize},
			hline{1,2}={1-Z}{\tablelinewidth},
			stretch=\tablestretch
		}
		Name & Processor & Architecture & Cores & Memory \\
		
		lin\_\#\# &
		{Intel\superR Xeon\superR Gold 6140 \\ CPU @ 2.30GHz} &
		x86-64 &
		18 &
		16 GB \\
		
		mac\_\#\# &
		Apple M3 &
		ARM64 &
		8 &
		16 GB \\
		
		win\_\#\# &
		{Intel\superR Core\superTM i5-7400 \\ CPU @ 3.00GHz} &
		x86-64 &
		4 &
		24 GB \\
	\end{tblr}
	
	\vspace*{0.4cm}
	\begin{tblr}{
			colspec={
				Q[c,m,1.5cm]
				Q[c,m,7cm]},
			rows={font=\footnotesize},
			row{1}={abovesep+=2pt, font=\normalsize},
			hline{1,2}={1-Z}{\tablelinewidth},
			stretch=\tablestretch
		}
		Name & OS / Kernel \\
		
		lin\_\#\# &
		Linux, kernel 4.18.0-553.el8\_10.x86\_64 \\
		
		mac\_\#\# &
		macOS Sequoia 15.6.1 \\
		
		win\_\#\# &
		Windows 10 Professional 22H2 19045.6282 \\
	\end{tblr}
	
	\vspace*{0.4cm}
	\begin{tblr}{
			colspec={
				Q[c,m,1.5cm]
				Q[c,m,7cm]},
			rows={font=\footnotesize},
			row{1}={abovesep+=2pt, font=\normalsize},
			hline{1,2}={1-Z}{\tablelinewidth},
			stretch=\tablestretch
		}
		Name & Compiler \\
		
		\#\#\_clang &
		Clang 21.1.0 \\
		
		\#\#\_gcc &
		GCC 14.3.0 \\
		
		\#\#\_intel &
		Intel 2025.2 (ICC) \\
		
		\#\#\_msvc &
		MSVC 2022 Community Version 19.44.35215 for x64 \\
	\end{tblr}
	\caption{
		Configurations of benchmark
		---
		Six configurations were tested: \config{lin}{gcc}, \config{mac}{clang},
		\config{mac}{gcc}, \config{win}{gcc}, \config{win}{intel}, 
		\config{win}{msvc}.
	}
	\label{tab:configurations}
\end{table}

\subsection{Benchmark}

The benchmark consisted in a home-made framework based on a unique main file,
in which all scenarios were built, but only one was selected and run. For each
scenario, to ensure the absence of device speed reduction due to an unforeseen
background process, the plain and intrinsic versions were alternately run with
the following scheme \code{[[plain,intrinsic],} \code{[plain,intrinsic],}
\code{...]} with size 50. Each time, their results were checked to be equal and
their respective execution times were measured. As a matter of precaution, the
scheme \code{[[plain,plain,...],} \code{[intrinsic,intrinsic,...]]} with
sub-sizes 50 were also tested, but similar execution times were obtained. The
resulting execution times were reduced to a pair of mean / standard deviation,
one for each code version. The execution time improvement / degradation was
evaluated through the execution time ratio of intrinsic version over plain
version, defined by:
\begin{equation}
	\ETR = \frac{\ETI}{\ETP}
	\label{equ:execution-time-ratio}
\end{equation}
where $\ETI$ and $\ETP$ are respectively the intrinsic and the plain
execution times. It gives a positive value, where a value below one (or 100\%)
expresses an execution time reduction of intrinsic version, and reciprocally
for a value above one (or 100\%). In our particular case, the lower the value of
$\ETR$, the better. The standard deviation of the execution time ratio was
computed through the equation:
\begin{equation}
	\stdf(\ETR) =
		\left\lvert \frac{\ETI}{\ETP} \right\lvert
		\sqrt{
			\left(\frac{\stdf(\ETI)}{\ETI}\right)^2
			+ \left(\frac{\stdf(\ETP)}{\ETP}\right)^2
		}
	\label{equ:standard-deviation-execution-time-ratio}
\end{equation}
where $\stdf(y)$ represents the standard deviation of variable $y$, and
variables $\ETI$ and $\ETP$ were supposed uncorrelated.

\section{Results}

\subsection{Execution time ratio}

\begin{figure}[htbp!]
	\vspace{-10pt}
	\includegraphics[width=\textwidth]{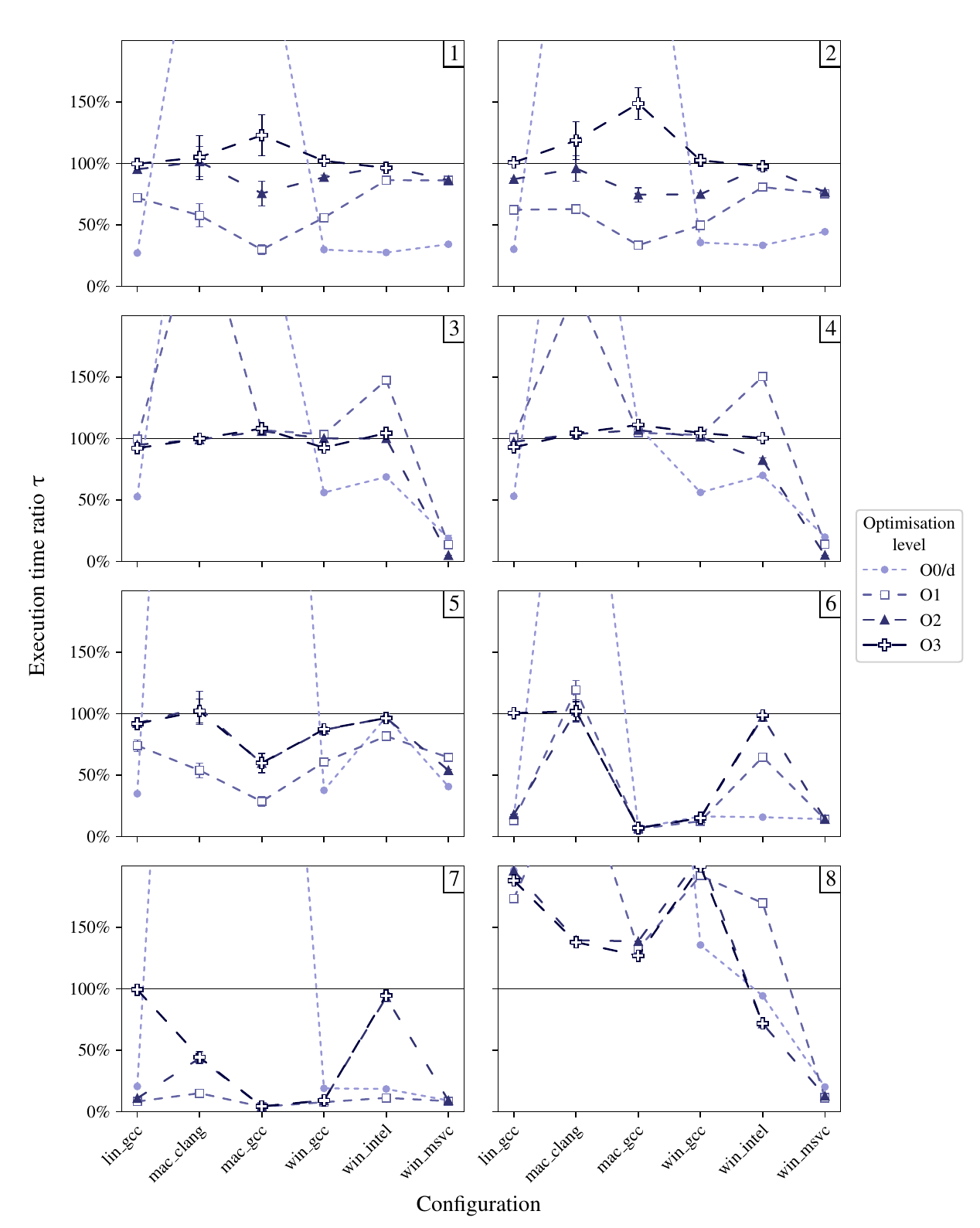}
	\caption{
		Benchmarking results
		---
		Execution time ratio, based on \cref{equ:execution-time-ratio}, as a
		function of configuration and optimisation level. Standard deviation is 
		based on \cref{equ:standard-deviation-execution-time-ratio}. Each 
		graph corresponds to a scenario (see \cref{list:plain-scenarios} for 
		details), whose index is indicated in the top right corner.
	}
	\label{fig:curve-graph}
\end{figure}

As a first set of results, \cref{fig:curve-graph} displays the execution time
ratio as a function of configurations and optimisation level, for each scenario.
All graphs show the range going from 0\% to 200\%. A first observation is the
drastic execution time ratio that appears on macOS system for optimisation level
O0, systematically for Clang compiler and depending on scenario for GCC
compiler. On these configurations and with deactivated optimisation, the use of
intrinsics was clearly destructive. For example, for optimisation level O0 and 
\config{mac}{clang}, the execution time ratio varies between 320\% and 910\%
depending on the scenario. Another interesting point appears in the four first
scenarios (top graphs in \cref{fig:curve-graph}, from 1 to 4), where scenarios
pairs 1 / 2 and 3 / 4 are respectively quite similar. Consequently, the
situation with index offset seems equivalently managed by compilers, whatever
the configuration. For basic operations (graphs 1 and 2 in
\cref{fig:curve-graph}), the use of intrinsics leads to promising execution time
ratio when the optimisation level is set to O1 (square markers), reaching
\valperc{29.76}{3.8} for \config{mac}{gcc} configuration in scenario 1. However,
the interest is much less clear when optimisation level is set to most
aggressive (02 for MSVC++ and O3 for other compilers). Indeed, the execution
time ratio is much closer to 100\%, which means that no change was brought by
intrinsic version. The trend even reverses for \config{mac}{gcc} configuration,
where the execution time ratio is around \valperc{148.8}{12.9} in scenario 2.
This observation is similar for advanced operations (graphs 3 and 4 in
\cref{fig:curve-graph}), where the optimisation level O3 gives execution time
ratio even closer to 100\%. An exception occurs for \config{win}{msvc}
configuration where the optimisation level O2 generated an execution time ratio
of \valperc{5.0}{0.1} in scenario 3, which constitutes an impressive speedup.
When it comes to conditional scenarios (bottom graphs in \cref{fig:curve-graph},
from 5 to 8), the simple case of condition based on index (scenario 5) gives a
similar trend. A promising execution time ratio appears at low optimisation
level, but it is questioned when the latter is increased, especially for
\config{mac}{clang} and \config{mac}{intel} where the execution time ratio is
close to 100\%. This result was expected in the sense that the conditional
branches were made based on vector index, which could be predicted and
auto-vectorised by compilers. For this reason, in scenarios 6 to 8, the
conditions were based on random data so that compiler could not predict the
branch routing in advance for each vector index, making the auto-vectorisation
difficult. The consequence is clearly captured in scenario 6, where even at
optimisation level O3, \config{mac}{gcc} and \config{win}{gcc} present
impressive execution time ratios of \valperc{6.9}{1.0} and \valperc{15.0}{0.2}
respectively. Nevertheless, unexpectedly, \config{lin}{gcc}, \config{mac}{clang}
and \config{win}{intel} still present execution time ratio close to 100\%. This
observation occurs once again in scenario 7 (with additional conditional
sub-branches), except for \config{mac}{clang} configuration that generates a
non-negligible execution time ratio of \valperc{44.3}{4.8} at optimisation level
O3. Finally, in scenario 8 (conditions with sub-branches and advanced
operations), the trend completely reverses: all configurations present
noticeable execution time ratios above 100\%, which signifies an important
increase of execution time due to intrinsic version. The only significant
configuration remains \config{win}{msvc}, that still present really promising
values, for example at optimisation level O2 where it equals
\valperc{12.8}{0.6}.

\subsection{Execution time}

\begin{figure}[htbp!]
	\vspace{-20pt}
	\includegraphics[width=\textwidth]{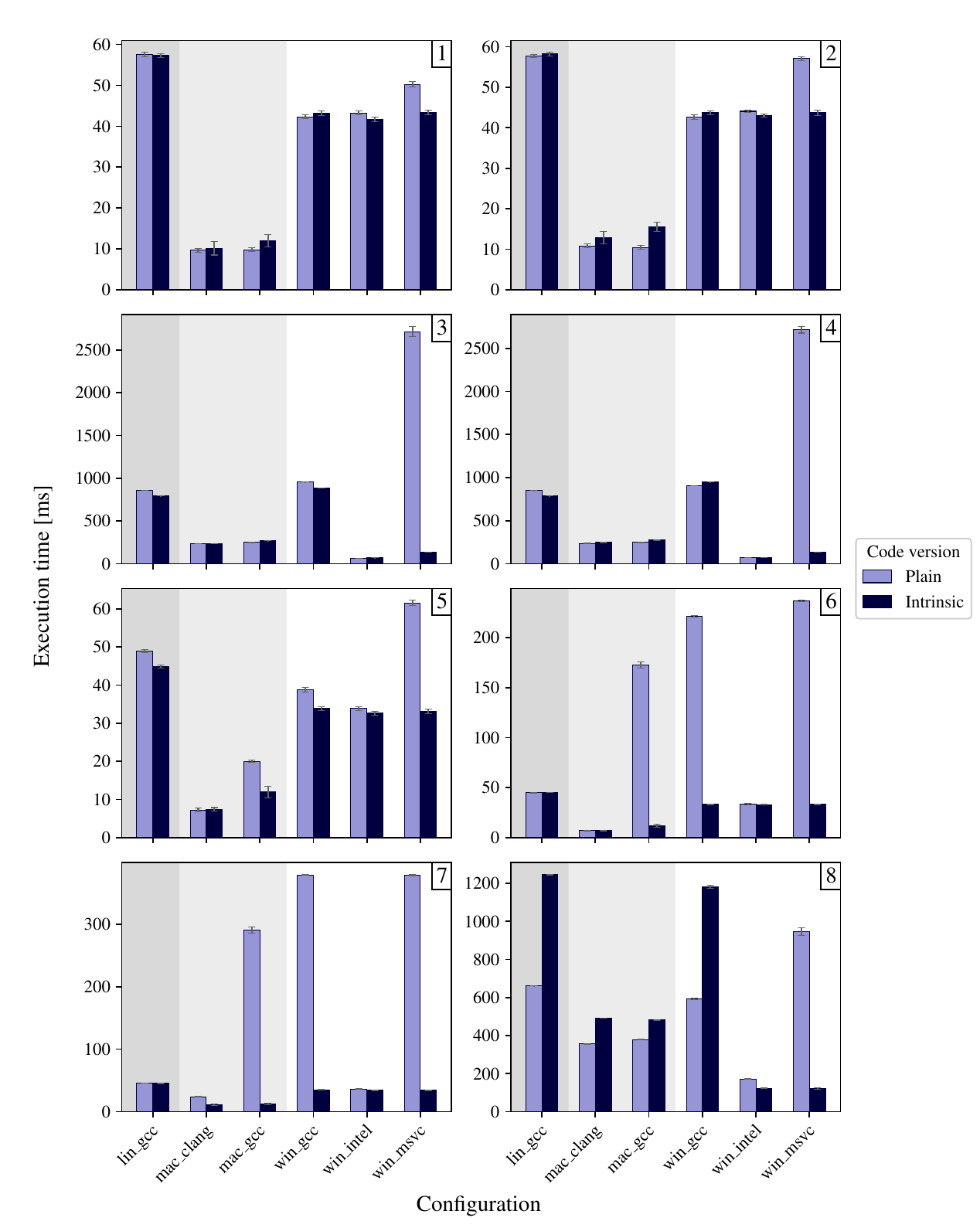}
	\caption{
		Benchmarking results
		---
		Execution times ($\ETP$ and $\ETI$) as a function of configuration, for 
		most aggressive optimisation (O2 for MSVC++ and O3 for other 
		compilers). Standard deviation is the sample uncertainty (fifty runs). 
		Each graph corresponds to a scenario (see \cref{list:plain-scenarios} 
		for scenarios details), whose index is indicated on the top right 
		corner. Background areas distinguish each device (Linux, macOS or 
		Windows) from one another.
	}
	\label{fig:bar-graph}
\end{figure}

A major conclusion of execution time ratio analysis is that, globally, the
\config{win}{msvc} is notably better performing when the intrinsic version is
used. However, the execution time ratio defined by
\cref{equ:execution-time-ratio} introduces a major drawback: it does not reveal
the order of magnitude of execution time. This is useful to make a comparison
between all configurations, but it fails to answer an important question:
\textit{For a particular device, which configuration is the best?} In this
perspective, \cref{fig:bar-graph} displays the execution times ($\ETP$ and
$\ETI$) for the most aggressive optimisation levels (O2 for MSVC++ and O3 for
other compilers). Background areas were added to distinguish each device (Linux,
macOS or Windows) from one another. Thus, precisely comparing the absolute
execution times from different areas is not relevant, as it compares different
devices that certainly perform differently. Still, it is possible to compare the
trends. Scenarios with basics operations only (scenarios 1 and 2 in
\cref{fig:bar-graph}), including conditions on vector index (scenario 5 in
\cref{fig:bar-graph}) are quite stable and cheap, plain and intrinsic versions
are clearly comparable. At the opposite, the introduction of advanced operations
(\code{cos}, \code{sqrt}, \code{pow}, etc., in scenarios 3 and 4 in
\cref{fig:bar-graph}) drastically increases the execution time, especially for
\config{win}{msvc} configuration. Similarly, the conditions on random data
(scenarios 6, 7 and 8) gradually increases the execution time, especially as the
program is complicated. These observations remain predictable and do not
constitute the most interesting aspects of \cref{fig:bar-graph}. The most
noticeable result is the case of \config{win}{intel} / \config{win}{msvc} pair.
In every scenario, \config{win}{msvc} sees an important speedup thanks to
intrinsic version, but it systematically makes the execution time reach the
order of magnitude of \config{win}{intel}. In other words, \config{win}{msvc}
configuration requires the use of intrinsics to reach the performance that
\config{win}{intel} already has in plain version. For scenarios 6 and 7 in
\cref{fig:bar-graph}, \config{win}{gcc} behaves the same. For example, in
scenario 6 (simple operations on random data with basic operations),
\config{win}{gcc} and \config{win}{msvc} respectively goes from
\valms{221.4}{0.6} to \valms{33.3}{0.5} and from \valms{236.5}{0.7} to
\valms{33.2}{0.5} thanks to intrinsic version, but \config{win}{intel} is
already at \valperc{33.6}{0.5} with plain version. A similar pair exists on
macOS device, where \config{mac}{gcc} systematically reaches back the execution
time of \config{mac}{clang} when the intrinsic version is used, particularly for
scenarios 6 and 7 in \cref{fig:bar-graph}.

\section{Discussion}

The results of the proposed cross-configuration benchmark demonstrates key
information: the interest of intrinsics is not only dependent on the program to
be vectorised, but also on the configuration (\iE OS, architecture and
compiler). It is difficult to state clear correlation, particularly concerning
the chip architecture. For example, in scenario 6 at optimisation level O3, both
\config{mac}{gcc} and \config{win}{gcc} configurations behave similarly whereas
they are run on two different architectures (respectively, ARM64 and x86-64). On
the opposite, for the same scenario and optimisation level, \config{lin}{gcc}
and \config{win}{gcc} behave quite differently whereas they run on the same
architecture. When it comes to compilers, some trends come up from the results:
GCC is poorly improved by intrinsics on Linux, Clang performs better than GCC on
macOS, ICC performs better than GCC and MSVC++ on Windows and MSVC++ is highly
improved by intrinsics on Windows. Another noticeable information is the
instability of advanced operations (\code{cos}, \code{sqrt}, \code{pow}, etc.).
Even if their implementation in intrinsic version leads at best to an unchanged
execution time in scenarios 3 and 4, it still leads to an important execution
time increase in scenario 8, which is clearly counterproductive from a
development perspective. This constitutes a good example of the warning stated
by Intel\superR Intrinsics Guide \cite{intel_intrinsics_2024}. When looking at
the information of a SVML (Short Vector Math Library) function: it is tagged as
\isay{SEQUENCE}, which is described as \isay{This intrinsic generates a sequence
of instructions, which may perform worse than a native instruction.}. Thus, the
advanced operations must be addressed carefully with intrinsics. This fact is
already well addressed in literature, where efforts are deployed to propose
approximation functions of these operations by using only fundamental operations
(addition, multiplication, subtraction, division) \cite{kusaka_fast_2022,
stpiczynski_parallel_2024, goyal_novel_2025}.

\begin{figure}[b!]
	\tikzstyle{arrow} = [
	->,
	>=stealth,
	thick
	]
	\tikzstyle{line} = [
	-,
	thick
	]
	\tikzstyle{styleDevice}=[
	trapezium,
	draw,
	text centered,
	trapezium left angle=70,
	trapezium right angle=110,
	minimum height=2.8em,
	text width=2cm
	]
	\tikzstyle{styleContext} = [
	rectangle,
	draw,
	text centered,
	rounded corners,
	minimum height=2em,
	text width=1.5cm,
	fill=blue!20
	]
	\tikzstyle{styleSolution} = [
	rectangle,
	draw,
	text centered,
	minimum height=2em,
	fill=green!20,
	text width=2.1cm
	]
	\tikzstyle{styleStartDevice} = [
	rectangle,
	draw,
	text centered,
	rounded corners,
	minimum height=2em,
	text depth=-2pt,
	fill=blue!20
	]
	\tikzstyle{styleUnkown} = [
	rectangle,
	draw,
	text centered,
	fill=green!20
	]
	\begin{tikzpicture}
		\small
		\node (device)
		[styleStartDevice]
		{\textbf{Device}};
		\node (win)
		[styleDevice, below=0.5cm of device]
		{Windows (x86-64)};
		\node (lin)
		[styleDevice, right=0.55cm of win]
		{Linux (x86-64)};
		\node (mac)
		[styleDevice, right=0.55cm of lin]
		{macOS (ARM64)};
		\node (otherDevice)
		[styleDevice, right=0.55cm of mac]
		{Other \\ device};
		
		\node (decWinICC)
		[styleContext, below=3.8cm of win]
		{ICC available?};
		\node (decWinMSVC)
		[styleContext, right=0.8cm of decWinICC]
		{MSVC++ available?};
		\node (decWinGCC)
		[styleContext, right=0.8cm of decWinMSVC]
		{GCC available?};
		\node (decLinGCC)
		[styleContext, below=0.5cm of lin]
		{GCC available?};
		\node (decMacClang)
		[styleContext, below=0.5cm of mac]
		{Clang available?};
		\node (decMacGCC)
		[styleContext, right=0.8cm of decMacClang]
		{GCC available?};
		
		\node (solWinICC)
		[styleSolution, below=0.6cm of decWinICC]
		{Use ICC, \\ without intrinsic.};
		\node (solWinMSVC)
		[styleSolution, anchor=base]
		at (decWinMSVC |- solWinICC.base)
		{Use MSVC++, \\ with intrinsics wherever it is needed.};
		\node (solWinGCC)
		[styleSolution, anchor=base]
		at (decWinGCC |- solWinMSVC.base)
		{Use GCC, \\ with intrinsics for condition branches.};
		\node (solLin)
		[styleSolution, below=0.6cm of decLinGCC]
		{Use GCC, \\ without intrinsic.};
		\node (solMacClang)
		[styleSolution, anchor=base]
		at (decMacClang |- solLin.base)
		{Use Clang, \\ without intrinsic.};
		\node (solMacGCC)
		[styleSolution, anchor=base]
		at (decMacGCC |- solMacClang.base)
		{Use GCC, \\ with intrinsics for condition branches.};
		
		\node (otherWin)
		[styleUnkown, right=0.8 of decWinGCC]
		{?};
		\node (otherLin)
		[styleUnkown, right=0.8 of decLinGCC]
		{?};
		\node (otherMac)
		[styleUnkown, right=0.8 of decMacGCC]
		{?};
		\node (otherOtherDevice)
		[styleUnkown, right=0.8 of otherDevice]
		{?};
		
		\draw [arrow] (device.south)
		-- ++(0,-0.2) coordinate (div)
		-| (otherDevice.north);
		\draw [arrow]
		let \p1=(div), \p2=(win.north) in (\x2,\y1)
		-- (win.north);
		\draw [arrow]
		let \p1=(div), \p2=(lin.north) in (\x2,\y1)
		-- (lin.north);
		\draw [arrow]
		let \p1=(div), \p2=(mac.north) in (\x2,\y1)
		-- (mac.north);
		
		\draw [arrow] (win) -- (decWinICC);
		\draw [arrow] (decWinICC) -- node[anchor=west] {Yes} (solWinICC);
		\draw [arrow] (decWinICC) -- node[anchor=south] {No} (decWinMSVC);
		\draw [arrow] (decWinMSVC) -- node[anchor=west] {Yes} (solWinMSVC);
		\draw [arrow] (decWinMSVC) -- node[anchor=south] {No} (decWinGCC);
		\draw [arrow] (decWinGCC) -- node[anchor=west] {Yes} (solWinGCC);
		\draw [arrow] (decWinGCC) -- node[anchor=south] {No} (otherWin);
		\draw [arrow] (lin) -- (decLinGCC);
		\draw [arrow] (decLinGCC) -- node[anchor=west] {Yes} (solLin);
		\draw [arrow] (decLinGCC) -- node[anchor=south] {No} (otherLin);
		\draw [arrow] (mac) -- (decMacClang);
		\draw [arrow] (decMacClang) -- node[anchor=west] {Yes} (solMacClang);
		\draw [arrow] (decMacClang) -- node[anchor=south] {No} (decMacGCC);
		\draw [arrow] (decMacGCC) -- node[anchor=west] {Yes} (solMacGCC);
		\draw [arrow] (decMacGCC) -- node[anchor=south] {No} (otherMac);
		\draw [arrow] (otherDevice) -- (otherOtherDevice);
	\end{tikzpicture}
	\caption{
		Flowchart for choice of intrinsics use, deduced from benchmark results. 
		The use of intrinsics is supposed available in the development 
		context (in terms of readability and maintainability). The objective of 
		such a routine is to optimise final performance, so by setting the 
		optimisation level to most aggressive (O2 for MSVC++ and O3 for other 
		compilers).
	}
	\label{fig:flowchart}
\end{figure}
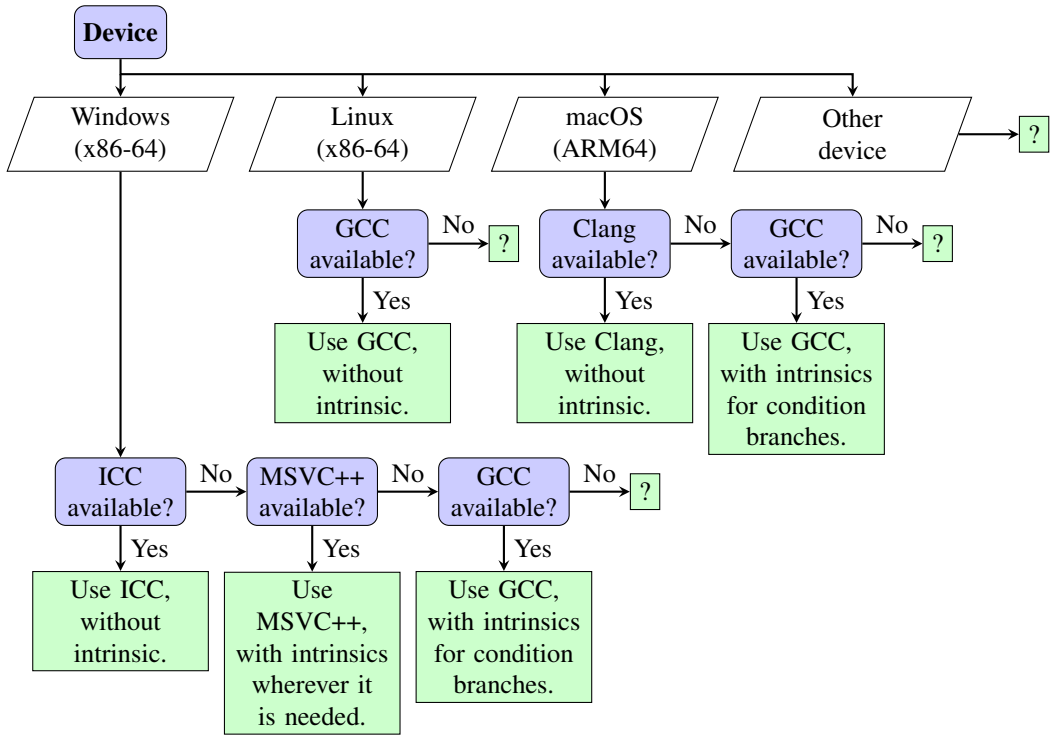

These results are interesting in the fact that they question the manner of
thinking programming, especially for compiled language such as C++: beyond the
implementation itself, the choice of the compiler is decisive. In other words,
before thinking about CPU vectorisation programming, it is wiser to check if the
right compiler and right optimisation options are used to enhance the
performance. The implementation of intrinsics is expensive in terms of code
tediousness, so it must be considered parsimoniously, as a last resort. As a
recall, the aim of this paper was to provide a routine helping in the choice of
intrinsics use. To fulfil this objective, \cref{fig:flowchart} proposes a flow
chart of generic routine to choose the right compiler and intrinsics use.

As a matter of critic of the proposed routine, it is hard to estimate to what
extent the proposed benchmark in not exhaustive enough. With the grand variety
of OS and chips, numerous behaviours may have not been captured (AMD chips,
Intel chips on macOS, ICC compiler on Linux, etc.), embodied by the \say{?}
cases in \cref{fig:flowchart}. It is also clear that the out-of-context
measurement, on single pieces of program, does not necessarily reflect the
performance reality of the main program that contains them, as discussed by
Gottschlag \eA (2020a and 2020b) \cite{gottschlag_automatic_2020,
gottschlag_avx_2020}. As a result, the proposed routine in \cref{fig:flowchart}
remains a guide and does not prevent the second piece of advice of Intel\superR
Intrinsics Guide \cite{intel_intrinsics_2024}: \isay{Consider the performance
impact of this intrinsic.}. The approach of this paper was not to question, but
on the opposite to emphasise, what constitutes to the authors' mind both first
main rules of optimisation:

\begin{enumerate}
	\item DOINN rule (Don't Optimise If Not Necessary).
	\item MAAC rule (Measure At Any Cost).
\end{enumerate}

The DOINN rule is fundamental to keep in mind that optimisation comes after a
technical need, not the reverse. The MAAC rule is important to prevent technical
bias to implement destructive solution: the systematic measurement of the
proposed optimisation solution (through execution time, CPU clocks, etc.) is the
essential way to state a robust conclusion about its real pertinence.

\section{Conclusion}

A cross-configuration benchmark was proposed to estimate the configurations
where AVX / NEON intrinsic functions were susceptible to notably improve
performance of a program. It was composed of eight scenarios commonly met in
development practices and eligible for SIMD vectorisation. Several compilers /
OS / architectures were tested: GCC on Linux (x86-64), Clang and GCC on macOS
(ARM64), GCC, ICC and MSVC++ on Windows (x86-64).

Impressive execution time reduction was captured in some cases thank to
intrinsic version (down to 5\% of the plain version execution time), for MSVC++
with advanced operations (\code{cos}, \code{sqrt}, \code{pow}, etc.) or cases of
conditional branches. However, the efficiency of intrinsic versions was not
systematic, even leading to important execution time increase when advanced
operations were used with conditional branches. Some trends were captured
depending on compilers and OS, revealing that GCC is poorly improved by
intrinsics on Linux (x86-64), Clang is more efficient than GCC on macOS (ARM64),
ICC is more efficient than GCC and MSVC++ on Windows (x86-64), and MSVC++ is
highly improved by intrinsics. In addition, a routine was proposed to guide the
developer to choose the use or not of intrinsics, depending on the configuration
constraints.

This paper mostly encourages, before the use of intrinsic functions, to explore
all the most accessible optimisation techniques, starting with the choice of the
right compiler and the right optimisation options.

\newpage

\section*{Acknowledgments}

The authors acknowledge Isabelle d'Ast for her precious feedbacks and help in
making the benchmark harmonised between operating systems. They also acknowledge
the daily support of CERFACS teams through their enriching discussions.

\section*{Credit authorship contribution statement}

T.B. built and run the benchmark, designed the figures, drafted and wrote the
manuscript, T.B. and J.L. reviewed the benchmark, validated the results and
reviewed the manuscript.

\section*{Declaration of generative AI and AI-assisted technologies in the 
writing process}

During the preparation of this work, T.B. occasionally used
\isay{\href{https://www.deepl.com}{DeepL}} as a French-to-English translator for
unique words or short expressions (up to five words). T.B. also used
\isay{\href{https://languagetool.org}{LanguageTool}} for grammar / spelling
check. After using this tool, the authors reviewed and edited the content as
needed and take full responsibility for the content of the published article.

\section*{Declaration of competing interest}

The authors declare no conflicts of interest regarding this manuscript.

\section*{Data availability}

The data that support the findings of this study are openly available at GitLab 
repository \cite{gitlab_avx_neon}.

\printbibliography

\end{document}